# ASSESSING ULTRASONIC AND OPTICAL FLOW VELOCIMETRY IN A MILLIFLUIDIC DEVICE USING OIL-IN-WATER EMULSIONS AS BLOOD MIMICKING FLUID

Estelle LU[1, 2], Williams FLORES CISTERNAS[3], Héloïse UHL[1, 2], Alexandre CHARGUERAUD[1, 2], Quentin GRIMAL[4], Guillaume RENAUD[5], Jean-Gabriel MINONZIO[3], Jacques FATTACCIOLI[1, 2]

**AFFILIATIONS**

[1] Laboratoire P.A.S.T.E.U.R., Département de Chimie, École Normale Supérieure, PSL Research University, Sorbonne Université, CNRS, 75005 Paris, France

[2] Institut Pierre-Gilles de Gennes pour la Microfluidique, 75005 Paris, France

[3] Escuela de Ingeniería Informática, Center of Interdisciplinary Biomedical and Engineering Research for Health - MEDING , Universidad de Valparaíso, Valparaíso, Chile

[4] Laboratoire Imagerie Biomédicale, CNRS INSERM Sorbonne Université

[5] Department of Imaging Physics, Delft University of Technology, Delft, The Netherlands (G.R.)

Corresponding authors: jean-gabriel.minonzio@uv.cl and jacques.fattaccioli@ens.psl.eu

## ABSTRACT

Blood-mimicking fluids (BMFs) play a critical role in ultrasonic imaging and Doppler flow studies by replicating the physical and acoustic properties of blood. This study introduces a novel soybean oil-in-water emulsion as a BMF with particle size and deformability akin to red blood cells. Using a millifluidic device, we cross-validated flow profiles through both Doppler velocimetry and optical particle tracking, demonstrating compatibility with theoretical Poiseuille flow models. The millifluidic chip, fabricated via stereolithography, provided an optimized platform for dual optical and ultrasonic assessments. Results showed strong agreement between the two methods across a range of flow rates, affirming the suitability of the emulsion for velocimetry applications. Furthermore, the acoustic properties of soybean oil droplets support their potential as an echogenic and stable alternative to conventional BMFs.

## 1. Introduction

Blood mimicking fluids (BMFs) are synthetic substances designed to simulate the acoustic and physical properties of human blood for use in ultrasonic imaging and Doppler flow studies [1]. These fluids typically consist of a water-glycerol base with suspended particles, aiming to replicate blood's density, viscosity, and acoustic characteristics [2]. While BMFs offer significant advantages over using actual blood, such as reduced biohazard risks and improved stability, they face challenges in accurately replicating blood's complex behavior [3]. Traditional BMFs often use solid particles like nylon or polystyrene at low concentration to mimic blood ultrasonic properties [1]. However, these particles cannot be used at concentrations as high as red blood cells in blood, and their non-deformable nature makes it difficult to study flow in small capillaries [1]. These limitations highlight the need for new BMF formulations that can better address these constraints.

Oil-in-water emulsions present a promising alternative, as they can be engineered with a narrow size distribution closely matching that of red blood cells [4]. Various vegetable oils can be used to create these emulsions, potentially offering a new class of BMFs with improved properties [5]. In this paper, we demonstrate that soybean oil-in-water emulsions can serve as effective BMFs. We provide a cross-calibration of the flow profile in a 3D printed millifluidic channel using both light microscopy and Doppler measurement techniques. Furthermore, we show that these measurements can be used for non-destructive determination of millifluidic channel dimensions by analyzing flow velocity as a function of the suspension's flow rate.

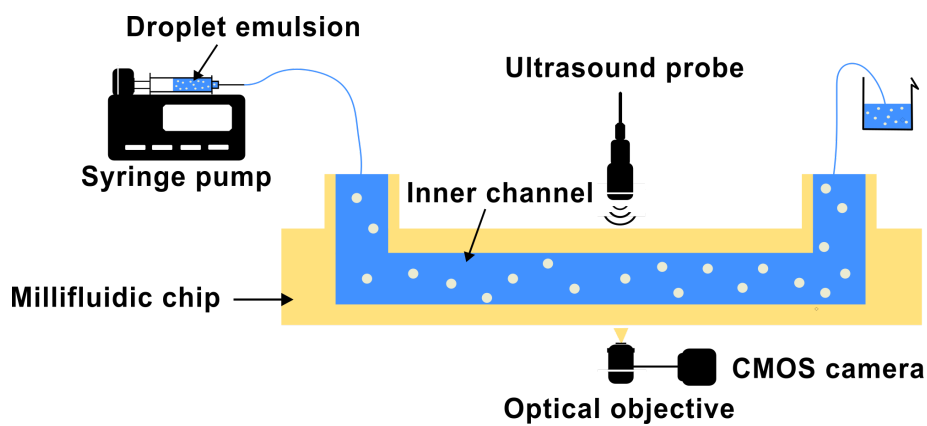

**Figure 1.** Schematic representation of the experimental setup for droplets velocity tracking through two techniques : optical tracking and ultrasound probing. Droplets are introduced, at a controlled flow rate, by a syringe pump through the millifluidic chip in which the velocities are measured.

## 2. Materials and Methods

### 2.1. Materials

Pluronic F-68 (Poloxamer 188, CAS no. 9003-11-6), Tween 20 (polyethylene glycol sorbitan monolaurate, CAS no. 9005-64-5) and soybean oil (CAS no. 8001-22-7) were purchased from Sigma-Aldrich (Saint Quentin Fallavier, France). Nano Clear resin (FunToDo) was purchased from Atome3D (Pechbonnieu, France). Propanol-2 (CAS no. 67-63-0) was purchased from VWR Chemicals (Rosny-sous-Bois, France). Ultrapure water (Millipore, 18.2 MΩ.cm) was used for all experiments. All reagents and materials were used as purchased, without any further purification.

### 2.2. Droplets Fabrication

Droplets were fabricated using a Shirasu Porous Glass membrane emulsification apparatus (SPG Technology Co., Ltd, Japan), with a porous membrane. This membrane is immersed in an aqueous solution containing 15% w/w of Pluronic F68 block copolymer. A pressure is applied by compressed air to the dispersed lipophilic phase, which is contained in a connected reservoir. Thus, this oil phase grows into droplets at the membrane pores (size = 2.1 µm). At a critical pressure, the droplets are extruded through the membrane and are stabilized in the continuous phase. The emulsion is continuously stirred throughout the process. The diameter dispersion was determined over 100 droplets and fitted with a Gaussian curve. The emulsion is stored in the dark at a constant temperature of 12°C in a Peltier-cooled cabinet. Prior to the experiments, the emulsion is diluted into either a 1% or a 10% particle volume fraction in the Pluronic F68 aqueous solution, for optical and ultrasonic measurements respectively.

## 2.3. Design and Fabrication of the Millifluidic Device

The millifluidic chip was designed using the CAD software Tinkercad. It features a closed channel positioned 1 mm from both the top and bottom surfaces, with a square cross-section of 2x2 mm². Inlet and outlet ports were incorporated on either side of the channel for connectivity. The STL chip design was sliced using Chitubox software. Fabrication was performed with an Elegoo Mars 2 Pro SLA 3D printer and a transparent FunToDo Nanoclear resin to obtain an optically transparent chip. A 26x76mm microscopy glass slide was affixed to the build platform with double-sided tape to ensure a smooth surface at the bottom of the chip. Printing parameters were adjusted to enhance adhesion between the resin and the glass slide. Particularly, exposure time for bottom layers was increased with respect to the default setting values. Both lifting speed and the delay at the platform's maximal height were increased with respect to the default setting values to ensure the clearest channel with the minimal amount of residual and uncured resin.

The printing process was paused after the last channel layers were completed, and 2-propanol was flushed through the channel to remove any residual resin. Printing was then resumed to complete the chip. The final object was detached from the glass slide, cleaned in a 2-propanol bath for 6 minutes, dried, and subjected to a final light curing process with the Elegoo Mercury Plus during 8 minutes.

## 2.4. Millifluidic Experiment

The inlet and outlet of the millifluidic chip were equipped with Luer locks and blunt-end Luer lock syringe needles. Plastic tubing (inner diameter: 0.02 inches, outer diameter: 1/16", Tygon, Saint-Gobain PPL Corp.) was attached to the opposite end of each needle. The inlet tubing was connected to a 5 mL plastic syringe (Terumo, Philippines) via an additional needle. The syringe, prefilled with a diluted droplet suspension, was linked through plastic tubing. All connections—including those at the inlet, outlet, and syringe—were sealed with Teflon tape to prevent leakage. Flow rates, ranging from 60 to 360 µL.min$^{-1}$, were controlled by a syringe pump (NE-4000, New Era Pump Systems, USA), which drove the droplets through the chip. The droplets were collected at the outlet in a glass bottle for reuse.

## 2.5. Flow Vector velocity mapping

A phased-array transducer (P4-1, ATL Philips, Bothell, WA, USA) operating at a center frequency of 2.5 MHz (96 elements, 0.295 mm pitch) was connected to a fully programmable ultrasound scanner (Vantage 256, Verasonics, Kirkland, WA, USA). A sequence of 15 tilted plane waves was transmitted with a time interval of 100 microseconds between each transmission. The last transmission is followed by a longer time interval, so that the frame rate (or temporal sampling rate of the flow) is 440 Hz. A continuous recording was acquired during 4.5 seconds. The steering angle of the transmit plane waves ranged -18 degrees to +18 degrees in the coupling gel. Images were reconstructed with a F-number of 1.3, therefore the spatial lateral resolution in the image was close to 1.1 mm (in the millifluidic phantom material). The temporal duration of the ultrasound pulse determines the spatial axial resolution (i.e., in the resolution along the depth direction), which was close to 0.7 mm (in the millifluidic phantom material). The height of the piezoelectric elements in the phased-array transducer determines the spatial elevational resolution (image thickness), which is close to 12 mm.

In order to extract the echo signal of the moving particles, we applied a temporal band-pass filter (4th order butterworth with cutoff frequencies 0.5 Hz and 6 Hz). The flow velocity and direction were estimated and mapped using a multi-angle plane wave imaging method [6] with refraction correction similar to [7]. The scanned region is described with two layers: the coupling gel and the phantom material. Applying an autofocusing method as described in [8] [9], we estimated the compressional wave speed in the coupling gel (1620 m.s$^{-1}$) and the millifluidic phantom material (2200 m.s$^{-1}$). The segmentation of the interface between the coupling gel and the phantom enabled the correction of wave refraction during image reconstruction and flow quantification. In order to enhance the signal-to-noise ratio and the spatial specificity of the signal received from the flowing particles, 3 synthetically focused transmit angles (-15 degrees, 0 degree and +15 degrees) were generated from the 15 plane wave transmit angles. Next, the ultrasound images were reconstructed with 4 receive angles (-20 degrees, -10 degrees, +10 degrees and +20 degrees) and wave refraction was taken into account. Therefore 12 combinations of transmit and receive angles were used to calculate the least-square estimate and map the velocity vector in the channel.

### 2.6. Optical Measurements

Brightfield microscopy images used to monitor droplet flow (1% oil volume fraction) were captured using a Leica DM IL LED inverted microscope equipped with 10× (N.A. = 0.22) and 20× (N.A. = 0.3) objectives. For droplet size measurements, images were acquired with a 40× objective (3-N-Achroplan, Zeiss, N.A. = 0.65). All images were recorded with an IDS camera (U3-3080CP-C-HQ, Rev.2.2), operated via the IDS Peak Cockpit software (version 1.5.0.0), which facilitated data retrieval, including corresponding timestamps for velocity measurements.
Images of the printed millifluidic device were captured using a motorized 2D/3D microscope (MRCL700 3D Imager Pro, MicroQubic AG, Switzerland), integrated with its proprietary camera.

### 2.7. Surface Analysis

The surface imaging was realized with the MicroCubic MRCL700 3D Imager Pro (MicroQubic AG, Switzerland). The surface profilometry was executed with the Dektak 6M Stylus Profiler (Veeco, Plainview, NY, USA), with a 12.5 µm radius stylus.

### 2.8. Image Analysis

Image analysis was performed with ImageJ/Fiji (version v1.54h) [10] and data analyses were performed with Matlab (MathWorks, Natick, MA, USA) software (R2022b version).

### 3. Results and discussion

We prepared a blood-mimicking fluid using a soybean oil-in-water emulsion suspension. While maintaining a continuous controlled flow rate of the suspension in a millifluidic channel, we measured the flow rate of the oil droplets both by optical particle trapping and Doppler velocimetry, as sketched in **Fig. 1**. Experimental details are given in the **Methods** section.

### 3.1. Formulation of the Blood-Mimicking Fluid

The particle suspension is an emulsion of soybean oil-in-water droplets fabricated using a membrane-emulsification apparatus, as sketched in **Figure 2A, B** and detailed in the **Methods** section. This technique consists in forcing an oil phase, here soybean oil, through a porous membrane into an aqueous phase containing surfactants, resulting in a size-controlled emulsion.

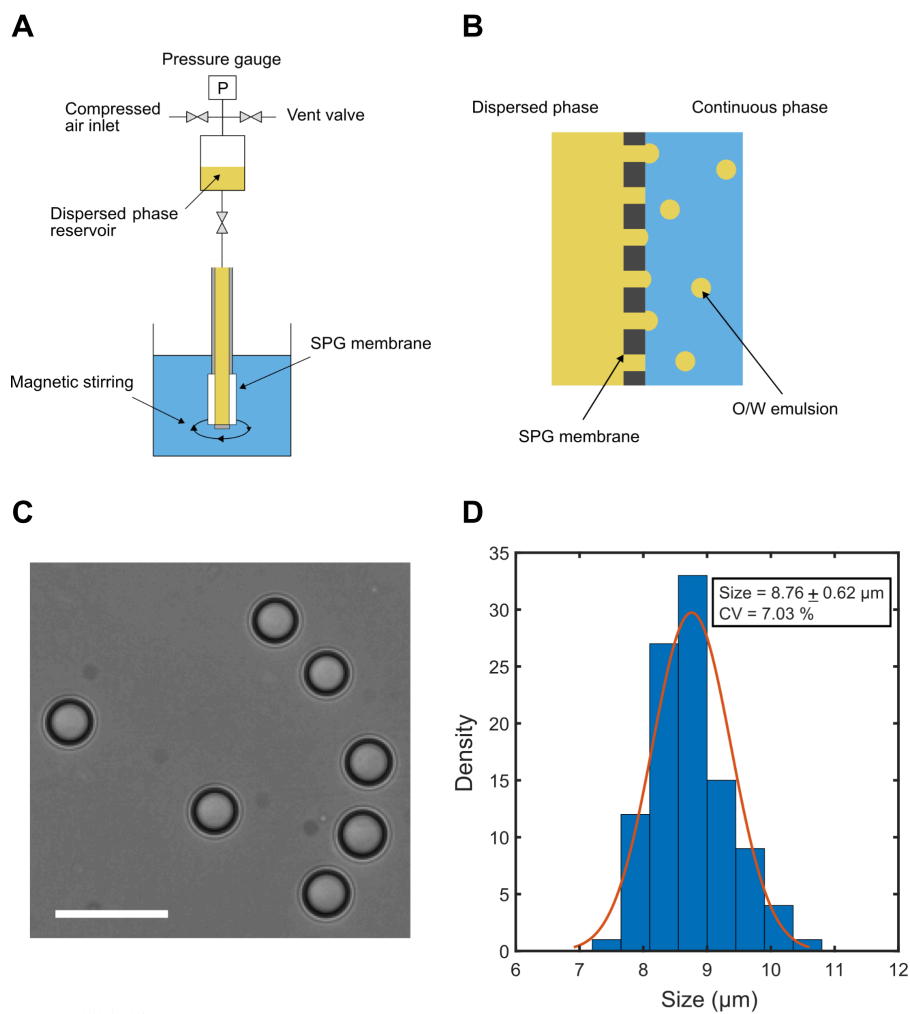

**Figure 2.** (A) Principle of the emulsification through a SPG membrane. (B) Porous membrane scheme for an oil-in-water emulsion. Scale bar : 20 µm (C) Brightfield image of soybean oil droplets used as blood-mimicking fluid for flow tracking. (D) Size distribution histogram of the oil-in-water emulsion.

The droplets are stabilized in an aqueous continuous phase of Pluronic F68 15% w/w (**Figure 2C**). In order to mimic the size of red blood cells, whose diameter is between 6 and 8 µm [11], we chose to work with a 2.1-µm-pore-size membrane, resulting in 8.8 $\pm$ 0.6 µm droplets, giving a variation coefficient of ca. 7% (**Figure 2D**). Prior to the experiments, the initial emulsion fabrication batch is diluted in Pluronic F68 aqueous solution at two different concentrations for the two different tracking strategies.

First, a droplet suspension with a volume fraction of 1% is prepared in order to track particles individually by optical method. Indeed, soybean oil has an optical refractive index of 1.48 [12], much higher than the refractive index of the aqueous continuous phase ($n$ = 1.33). This optical property gives an excellent optical contrast on the pictures of the droplets, hence easing the tracking if the droplet concentration is low enough to segment individual droplets.

Conversely, the acoustic velocity in vegetable oils, including soybean oil, in the ultrasonic range of 2 to 3 MHz is approximately 1460-1470 m.s$^{-1}$ at room temperature [13]. This value is very close to the acoustic velocity in water, which is around 1480 m.s$^{-1}$ in the same frequency range. Likewise, the mass density of soybean oil, ranging from 0.919 to 0.925 g.cm$^{-3}$, is close to the water value (1 g.cm$^{-3}$). These small velocity and density differences imply that the impedance contrast and therefore the amplitude of the backscattered signal are low (ref ?)   Hence, the ultrasonic strategy demanded a higher volume fraction to amplify the signals received from the sample. Therefore, we worked at a concentration of 10% in volume for the Doppler velocimetry.

### 3.2. Millifluidic Device Design and Fabrication

To perform the dual optical and ultrasonic flow velocity measurements, we designed a simple millifluidic chip consisting of a straight channel with a 2x2 mm square cross-section (**Figure 3A).** This model was manufactured by stereolithography (SLA), with a transparent resin. When comparing the channel's designed cross-sectional dimensions with those of the actual printed device, a discrepancy is observed, particularly in the height, which results in a 2x1.45 mm$^2$ rectangular cross-section instead of the intended square one, as shown on **Figure 3B**.

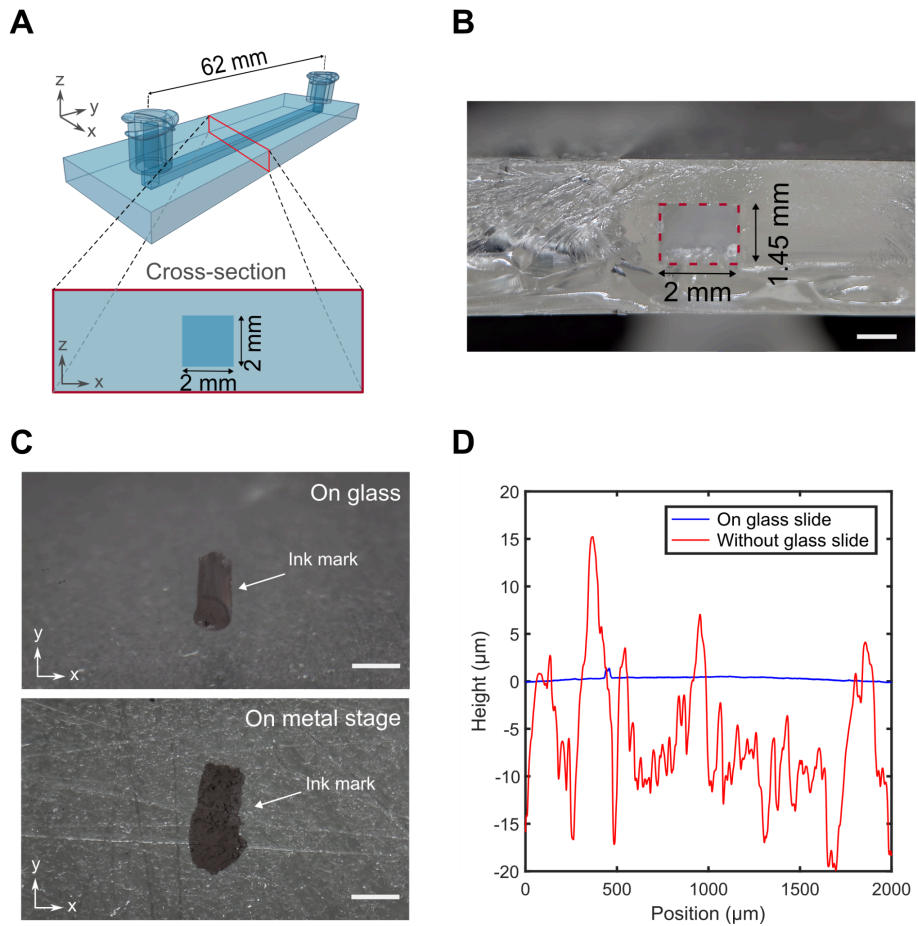

**Figure 3** (A) CAD representation of the printed millifluidic device with the cross sectional representation. (B) Cross sectional image of the printed device, with the resulting rectangular channel within the dotted rectangle. Scale bar : 2 mm. (C) Surfaces imaging of the device's bottom with and without a glass slide as the build plate. Scale bar : 1 mm. (D) Corresponding surface profilometry of the two printing techniques presented in (C).

The chip was printed using a glass slide as a building support using a custom-made setup (see **Methods** for details) to obtain a smooth bottom surface. This smoothness improvement is confirmed by comparing devices printed on the conventional metallic building platform with those printed on the glass slide. Profilometry analysis demonstrates a significant reduction in surface roughness when the glass slide is used. This ultimately improves the microscopic and ultrasonic imaging, by minimizing as much as possible light and ultrasound scattering at the chirp interface (see **Figure 3C, D**).

## 3.3. Optical and Ultrasonic Dual Velocity Measurements

To vary the flow rates within the 3D-printed microchannel, the continuous flow was adjusted across a range of 60 to 360 µL.min$^{-1}$ with a syringe pump, simulating the lower values commonly observed in the bloodstream [7,14].

For the optical characterization of the flow, the microscope objective is moved from the bottom to the top of the channel to capture the complete variation of the axial velocity profile. The channel's bottom is identified by observing adsorbed impurities on the inner wall of the bottom part of the channel. For each height increment of roughly 0.14 mm, an image sequence is recorded with a sufficiently high framerate (10 - 30 fps) to track the displacement of several particles within the observation field. For each time lapse recording, the instantaneous velocity V(z) of the particles is then measured by measuring their axial displacement Δx, measured by correlation analysis of pairs of successive images, over the sampling timeframe Δt recorded by the camera (**Figure 4A**). We can then reconstruct the parabolic velocity profile of the particles shown in **Figure 4B**, typical of a Poiseuille flow. This is in accordance with the value of the Reynolds number of the system, which remains small with our flow rate conditions.

Two-dimensional vector velocimetry was performed. Because the flow channel is parallel to the transducer array surface, the particle velocity corresponds to the lateral component of the estimated velocity vector. **Figure 4C** shows an example map of the particle velocity in the flow channel, as estimated with ultrasound imaging. As expected for a laminar flow, maximum flow velocity is observed in the middle of the channel (for a depth of approximately 5.5 mm). **Figure 4D** displays the estimated flow velocity as a function of time, at the position depicted in **Figure 4C** by the red cross. The lateral component of the estimated velocity vector was averaged temporally (over 4.5 seconds) and spatially within the flow channel in the ultrasound image.

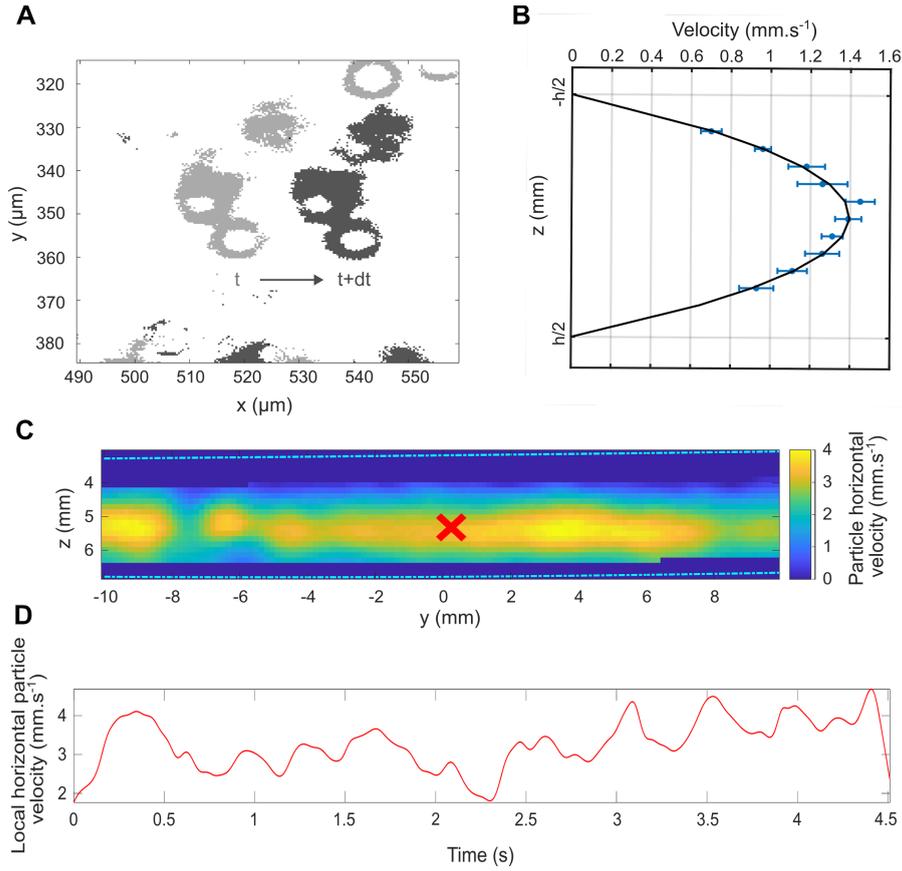

**Figure 4.** (A) Two consecutives images superposed representing a droplet displacement in the millifluidic chip used for optical tracking. This experiment was taken with a flow rate of 60 µL.min$^{-1}$, at a particle concentration of 1 vol%. (B) Velocity profile of droplets along the z-axis within the channel. The black line corresponds to the parabolic extrapolation coming from the optical measures. The resulting maximal velocity of 1.4 mm.s$^{-1}$ is deduced from the extrapolation. This profile was analyzed for a flow rate of 120 µL.min$^{-1}$. (C) Mapping of the flow velocity within the channel estimate with ultrasound flow vector velocimetry at a flow rate of 360 µL.min$^{-1}$. The red cross represents the local point where the velocity is measured as a function of time in (D).

### 3.4. Comparison of the Two Measurement Techniques

In the case of a Poiseuille flow, the maximal flow velocity writes as :

$$v_{max} = \frac{h^2}{8\mu L}\Delta P \qquad \text{(Equation 1)}$$

where ΔP is the pressure drop between the two extremities of the channel, µ is the dynamic viscosity of the emulsion suspension, w and h are respectively the channel's width and height, and L represents the length of the channel [15].

From the above maximum velocity , we then calculate its corresponding average flow velocity

$$V_{average} = \frac{h^é}{12\mu L}\Delta P = \frac{2}{3}v_{max} \qquad \text{(Equation 2)}$$

and the volumetric flow rate Q :

$$Q = \frac{wh^3}{12\mu L}\Delta P \qquad \text{(Equation 3)}$$

Finally, the average flow velocity V$_{average}$ can be expressed as a function of Q by combining Equations 2 and 3 :

$$V_{average} = \frac{1}{wh}Q \qquad \text{(Equation 4)}$$

**Figure 5** represents the average droplets velocity acquired optically (blue) and by Doppler technique (red). These two techniques are in good agreement for the different imposed flow rates. Additionally, the theoretical slope deduced from **Equation 4** and adjusted with the real dimensions of the 3D printed channel is drawn on the same figure. Figure 5 shows that both experimental results are in excellent agreement with theoretical values derived from a simple Poiseuille model. This analysis is not hindered by the two different emulsion volume fractions of 1% and 10%. It has been shown that at low concentrations such as those used in this study, the emulsion viscosity differs little [16] [17].

Blood-mimicking fluids (BMFs) are widely used and typically designed to meet the International Electrotechnical Commission (IEC) physical specifications, focusing on recreating artificial blood properties such as blood cell size. The BMF developed in this work achieves a controlled size distribution, with droplet diameters approximating those of red blood cells (~8 µm). This range is narrower than that of commercially available BMFs made with nylon, polyamide, or polystyrene microparticles [1] [18] [19]. While other IEC-specified properties like viscosity or density were not prioritized in this study, as they were not essential for the optical or Doppler measurements, they could be explored further to enhance applicability.

BMFs are essential for replicating blood behavior, yet their suitability for ultrasonic assessments can be limited, particularly due to low echogenicity issues [20]. In this study, the soybean oil-in-water emulsion demonstrated echogenic properties suitable for Doppler measurements. Edible oils, including soybean oil, are cheap, accessible, and have been studied for their ultrasonic properties, such as speed of sound characterization [21,22]. However, using soybean oil droplets for this purpose has not been extensively explored. Notably, this emulsion is highly stable, maintaining its properties for over a year, and exhibits deformability—an uncommon feature among conventional BMFs.

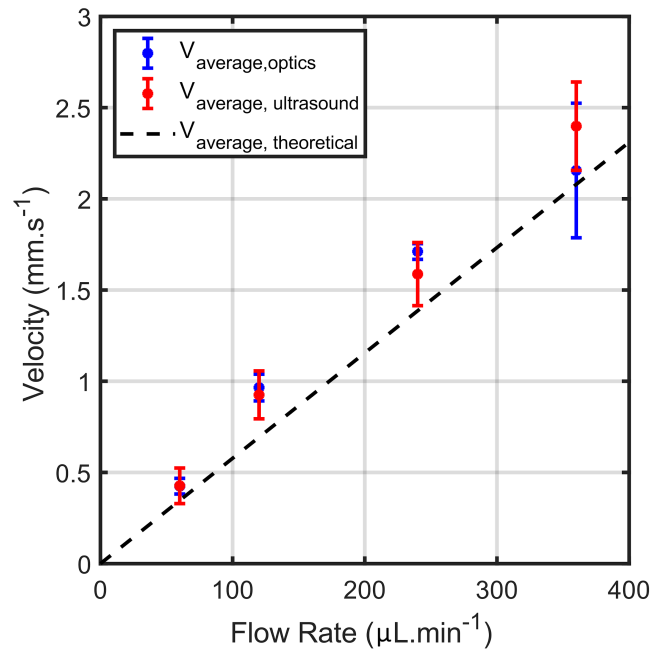

**Figure 5** Measurements of the average velocity of the droplet by optical microscopy (blue) and ultrasound probing (red) for various flow rates, along with the theoretical velocity deduced from the corrected geometrical dimensions. The dotted line represents the theoretical average velocity obtained taking *w* = 2 mm and *h* = 1.45 mm according to the measurements shown in **Fig. 3**.

Compared to polymers like nylon or polystyrene, soybean oil emulsions have a lower acoustic velocity (~1460-1470 m.s$^{-1}$, closer to water), which can limit certain applications. However, their shear-thinning behavior better mimics blood's rheology [23], potentially allowing higher droplet concentrations (~40%) similar to red blood cells. This opens opportunities to explore flow measurements in narrow, capillary-like channels while adhering to acoustical method standards to minimize attenuation and scattering.

## 4. Conclusion

This study demonstrates the feasibility of using soybean oil-in-water emulsions as a blood-mimicking fluid for both optical and ultrasonic flow velocity measurements. The emulsion achieves a precise size distribution similar to red blood cells and exhibits excellent stability and realistic echogenicity. Experimental results confirmed strong agreement between optical and Doppler measurements, validating the emulsion's performance in a 3D-printed millifluidic device. While certain physical properties, such as viscosity, were not optimized, the current formulation proved effective for velocimetry applications. The deformability and acoustic properties of the droplets make them a promising candidate for future studies, particularly in mimicking complex blood behavior. Further research could expand their utility by tailoring rheological and acoustic properties to specific experimental needs.

**Acknowledgments**

This work benefited from the technical contribution of the joint service unit CNRS UAR 3750. The authors would like to thank the engineers of this unit (and in particular Audric Jan and Elian Martin) for their advice during the development of the experiments. We thank Bastien Venzac (LAAS-CNRS) for additive manufacturing discussions. We thank Michel Cloître (3CM, ESPCI-PSL) for fruitful discussions about the viscosity of emulsion suspensions.

**Fundings**

This work was funded by ANID ECOS-Sud 20061 and ANID / Fondecyt / Regular 1241091.

**Author contribution**

JF, GR and JGM. designed the study. EL, WFC, QG and JGM conducted the experiments. Profilometry : AC. Droplet fabrication: HU. Millifluidic Device Design and Fabrication: EL, WFC, JF. EL, JF, GR, JGM analyzed the results and wrote the manuscript. All authors approved the final version of the article.

**Competing interests**

The authors declare that they have no competing interests.

**Data and materials availability**

All data needed to evaluate the conclusions in the paper are present in the papers. Additional data related to this paper may be requested from the authors.

**References**

[1]   K.V. Ramnarine, D.K. Nassiri, P.R. Hoskins, J. Lubbers, Validation of a new blood-mimicking fluid for use in Doppler flow test objects., Ultrasound Med. Biol. 24 (1998) 451–459. https://doi.org/10.1016/s0301-5629(97)00277-9.

[2]   X. Zhou, P.R. Hoskins, Testing a new surfactant in a widely-used blood mimic for ultrasound flow imaging., Ultrasound. 25 (2017) 239–244. https://doi.org/10.1177/1742271X17733299.

[3]   A.A. Oglat, A Review of Blood-mimicking Fluid Properties Using Doppler Ultrasound Applications., J. Med. Ultrasound. 30 (2022) 251–256. https://doi.org/10.4103/jmu.jmu_60_22.

[4]   Y.N. Shariffa, T.B. Tan, U. Uthumporn, F. Abas, H. Mirhosseini, I.A. Nehdi, Y.H. Wang, C.P. Tan, Producing a lycopene nanodispersion: Formulation development and the effects of high pressure homogenization., Food Res. Int. 101 (2017) 165–172. https://doi.org/10.1016/j.foodres.2017.09.005.

[5]   S.M. Ali, B. Ali, Acoustics Impedance Studies in Some Commonly Used Edible Oils, (2014).

[6]   B.Y.S. Yiu, A.C.H. Yu, Least-Squares Multi-Angle Doppler Estimators for Plane-Wave Vector Flow Imaging., IEEE Trans. Ultrason. Ferroelectr. Freq. Control. 63 (2016) 1733–1744. https://doi.org/10.1109/TUFFC.2016.2582514.

[7]   S. Salles, J. Shepherd, H.J. Vos, G. Renaud, Revealing intraosseous blood flow in the human tibia with ultrasound., JBMR PLUS. 5 (2021) e10543. https://doi.org/10.1002/jbm4.10543.

[8]   G. Renaud, P. Kruizinga, D. Cassereau, P. Laugier, In vivo ultrasound imaging of the bone cortex., Phys. Med. Biol. 63 (2018) 125010. https://doi.org/10.1088/1361-6560/aac784.

[9]   R. Waasdorp, D. Maresca, G. Renaud, Assessing transducer parameters for accurate medium sound speed estimation and image reconstruction., IEEE Trans. Ultrason. Ferroelectr. Freq. Control. 71 (2024) 1233–1243. https://doi.org/10.1109/TUFFC.2024.3445131.

[10]  J. Schindelin, I. Arganda-Carreras, E. Frise, V. Kaynig, M. Longair, T. Pietzsch, S. Preibisch, C. Rueden, S. Saalfeld, B. Schmid, J.-Y. Tinevez, D.J. White, V. Hartenstein, K. Eliceiri, P. Tomancak, A. Cardona, Fiji: an open-source platform for biological-image analysis., Nat. Methods. 9 (2012) 676–682.

https://doi.org/10.1038/nmeth.2019.

[11] M. Diez-Silva, M. Dao, J. Han, C.-T. Lim, S. Suresh, Shape and Biomechanical Characteristics of Human Red Blood Cells in Health and Disease., MRS Bull. 35 (2010) 382–388. https://doi.org/10.1557/mrs2010.571.

[12] E.M. Abdo, O.E. Shaltout, H.M.M. Mansour, Natural antioxidants from agro-wastes enhanced the oxidative stability of soybean oil during deep-frying, LWT. 173 (2023) 114321. https://doi.org/10.1016/j.lwt.2022.114321.

[13] C. Javanaud, R.R. Rahalkar, Velocity of sound in vegetable oils, Fett/Lipid. 90 (1988) 73–75. https://doi.org/10.1002/lipi.19880900208.

[14] V. Grand-Perret, J.-R. Jacquet, I. Leguerney, B. Benatsou, J.-M. Grégoire, G. Willoquet, A. Bouakaz, N. Lassau, S. Pitre-Champagnat, A novel microflow phantom dedicated to ultrasound microvascular measurements., Ultrason Imaging. 40 (2018) 325–338. https://doi.org/10.1177/0161734618783975.

[15] H.A. Stone, Introduction to fluid dynamics for microfluidic flows, in: H. Lee, R.M. Westervelt, D. Ham (Eds.), CMOS Biotechnology, Springer US, Boston, MA, 2007: pp. 5–30. https://doi.org/10.1007/978-0-387-68913-5_2.

[16] V. Vand, Theory of viscosity of concentrated suspensions, Nature. 155 (1945) 364–365. https://doi.org/10.1038/155364b0.

[17] J.W. Bullard, A.T. Pauli, E.J. Garboczi, N.S. Martys, A comparison of viscosity-concentration relationships for emulsions., J. Colloid Interface Sci. 330 (2009) 186–193. https://doi.org/10.1016/j.jcis.2008.10.046.

[18] M.L. Thorne, T.L. Poepping, R.N. Rankin, D.A. Steinman, D.W. Holdsworth, Use of an ultrasound blood-mimicking fluid for Doppler investigations of turbulence in vitro., Ultrasound Med. Biol. 34 (2008) 1163–1173. https://doi.org/10.1016/j.ultrasmedbio.2007.12.014.

[19] T. Yoshida, K. Tanaka, K. Sato, T. Kondo, K. Yasukawa, N. Miyamoto, M. Taniguchi, Blood-mimicking fluid for the Doppler test objects of medical diagnostic instruments, in: 2012 IEEE International Ultrasonics Symposium, IEEE, 2012: pp. 1–4. https://doi.org/10.1109/ULTSYM.2012.0403.

[20] V.A. Kumar, A.N. Madhavanunni, S. Nivetha, M.R. Panicker, On the Echogenicity of Natural Starch-Based Blood Mimicking Fluids for Contrast Enhanced Ultrasound Imaging: Preliminary In-vitro Experiments, in: 2024 IEEE South Asian Ultrasonics Symposium (SAUS), IEEE, 2024: pp. 1–4. https://doi.org/10.1109/SAUS61785.2024.10563527.

[21] A. Jiménez, M. Rufo, J. Paniagua, A. González-Mohino, L.S. Olegario, New findings of edible oil characterization by ultrasonic parameters., Food Chem. 374 (2022) 131721. https://doi.org/10.1016/j.foodchem.2021.131721.

[22] P.A. Oliveira, R.M.B. Silva, G.C. Morais, A.V. Alvarenga, R.P.B.C.- Félix, Speed of sound as a function of temperature for ultrasonic propagation in soybean oil, J. Phys.: Conf. Ser. 733 (2016) 012040. https://doi.org/10.1088/1742-6596/733/1/012040.

[23] M.A. Abdelaziz, J.A. Díaz A., J.-L. Aider, D.J. Pine, D.G. Grier, M. Hoyos, Ultrasonic chaining of emulsion droplets, Phys. Rev. Research. 3 (2021) 043157. https://doi.org/10.1103/PhysRevResearch.3.043157.